\documentclass{article}

\usepackage{arxiv}

\usepackage[utf8]{inputenc}
\usepackage[T1]{fontenc}
\usepackage{hyperref}
\usepackage{url}
\usepackage{booktabs}
\usepackage{amsfonts}
\usepackage{nicefrac}
\usepackage{microtype}
\usepackage{graphicx}
\usepackage{amsmath,amssymb}
\usepackage{bm}
\usepackage{mathrsfs}
\usepackage{amsthm}
\usepackage{multirow}
\usepackage{float}
\usepackage{xcolor}

\newcommand{\Escale}{E_{\mathrm{scale}}}
\newcommand{\Ham}{\mathcal{H}}
\newcommand{\Mom}{\mathcal{M}}
\newcommand{\Idiag}{\mathbf{X}}
\newcommand{\avg}[1]{\left\langle #1 \right\rangle}
\newcommand{\rms}{\mathrm{RMS}}

\title{Reliable Weyl Diagnostics for an Inhomogeneous Universe in Numerical Relativity}

\author{
  Hassan Ugail \\
  Centre for Visual Computing and Intelligent Systems \\
  University of Bradford \\
  United Kingdom \\
   \\
}

\begin{document}
\maketitle

\begin{abstract}
The electric and magnetic parts of the Weyl tensor provide a geometric diagnostic for describing the local
gravitational structure of an inhomogeneous universe, and they allow us to distinguish tidal, gravitomagnetic,
radiative, and silent regimes. However, when these diagnostics are computed from numerical relativity data,
their interpretation depends on how well the data satisfy the Einstein constraint equations. A small
constraint error can create a spurious magnetic signal, especially near a silent regime, and it may therefore
be mistaken for genuine radiative structure. In this work, we treat the Hamiltonian and momentum constraint
residuals as reliability gates for the Weyl diagnostics. Rather than imposing a fixed tolerance, we measure
how a constraint error affects the corresponding Weyl classification, and we then calibrate the tolerance
required to achieve a chosen accuracy. Our noise-induced synthetic tests show that the Hamiltonian residual
determines the reliability of the electric Weyl diagnostic, whereas the momentum residual determines the
reliability of the magnetic Weyl diagnostic. Furthermore, a real numerical relativity initial-data slice
confirms that no universal tolerance is adequate. Thus, the calibration measures the slice-level sensitivity
of the diagnostics rather than the truncation error of an evolution.
\end{abstract}


\section{Introduction}
\label{sec:intro}

The characterisation of the local gravitational dynamics of an inhomogeneous universe has become a concrete
computational task rather than a purely formal one. As we know, the scalar averaging of the Einstein
equations shows that inhomogeneity enters the effective large-scale dynamics through backreaction terms, and
the sign and the magnitude of these terms depend on the averaging domain. This was first established for
irrotational dust and later for wider fluid classes~\cite{buch2000,buch2001}. The size of the associated
effects, and the sense in which a Newtonian description remains adequate, have also been examined within a
systematic weak-field framework~\cite{greenwald}. In parallel, full numerical relativity data has been applied
directly to cosmological spacetimes. It has been used to quantify the departures from the Friedmann model in
nonlinear settings, to follow the growth of structure, and to establish the feasibility and the validation of
the approach~\cite{giblin,bentivegna,macph2017,macph2019}. Weak-field relativistic simulation schemes have
developed alongside these fully nonlinear treatments~\cite{adamek}, and the field has now reached a stage of maturity~\cite{aurrek}. Hence, it is natural to ask not only how a cosmological spacetime evolves,
but also how its local geometric content should be read off and reported.

A recent phase-space framework addresses this question~\cite{ugail}. It summarises a solution, on a chosen
spatial domain and at a chosen time, through a vector of scalar invariants that separates the physical
mechanisms of departure from a Friedmann background. Such a description is unavoidably explicit about the
observer, the foliation, and the averaging domain, since the same spacetime admits different scalar summaries
under different choices. This dependence is intrinsic to relativistic averaging and to the interpretation of
gauge and frame in cosmological numerical relativity~\cite{buch2020,ijjas}. Two of the physical axes in this
description is derived from the Weyl tensor. Relative to a fundamental observer $u^{a}$, the Weyl tensor splits
into an electric part $E_{ab}$ and a magnetic part $B_{ab}$, and the resulting gravito-electromagnetic analogy
assigns to them the roles of the tidal field and the gravitomagnetic field~\cite{maartens,costa}. The
magnetic part carries particular physical weight. Its presence signals non-local, and possibly radiative,
gravitational structure, and its behaviour has been studied both in the expansion of discrete universes and
in the visualisation of the tidal and frame-dragging structure around dynamical strong-field
sources~\cite{clifton,owen}. It is exactly this content that has been mapped, cell by cell, on relativistic
cosmological data. There, the electric and magnetic Weyl fields distinguish the tidal from the gravitomagnetic
regimes during structure formation, and invariant post-processing implementations are now available for this
purpose~\cite{munozbruni,ebweyl}.

Any classification of this kind rests on an implied assumption, i.e., that the fields supplied to it satisfy the
Einstein constraints closely enough that the computed $E_{ab}$ and $B_{ab}$ carry the meaning the classifier
assigns to them. Numerically evolved data never satisfy the Hamiltonian and momentum constraints exactly, and
perturbatively or synthetically constructed data satisfy them only to the order of their construction. The
standard response in numerical relativity is to monitor the constraint residuals as a global measure of
solution quality, and to design formulations and gauges that keep those residuals bounded. This concern runs
from the early conformal reformulations through to the constraint-damped
systems~\cite{shibata,bssn,alcubierre,gundlach,alic}. However, this practice does not provide a local and
quantitative link between the size of a residual and the trustworthiness of a specific mechanism
classification. Neither does it state how small the residual must be before that classification can be trusted. A small constraint violation contaminates the Weyl scalars, and near a classification boundary - for
example, a nominally silent slice that acquires a small magnetic part - the contamination can invert the label
and present an apparent radiative signal that in fact originates in numerical error.

In this work, we address that missing link. We treat the two constraint-residual axes of the diagnostic vector
not as passive monitors, but as sector-specific reliability gates, and we make the threshold of each gate an
outcome of measurement rather than a fixed constant. A reliability gate is defined on the diagnostic vector so
that a mechanism classification is admitted only where the associated constraint residual falls below a
calibrated tolerance. This tolerance is set from the data. For a given solution, we measure the induced error
in the dominant Weyl axis as a function of the constraint residual, we fit it as a power law, and we invert
the fit to obtain the residual tolerance associated with a prescribed accuracy under the calibrated model.
Furthermore, we verify the whole construction with an independent implementation. The real-data computations
of the constraints and the Weyl scalars are carried out with the \textsc{aurel} package~\cite{aurel}, which
is the successor to the invariant characterisation implementation~\cite{ebweyl}. Hence, the reliability of the
framework is checked on the same machinery a practitioner would use, rather than re-derived. Note that the
real-data anchor injects a constraint-violating perturbation into a fixed initial-data slice and recomputes
the diagnostics. Thus, it calibrates the slice-level sensitivity of those diagnostics rather than the
truncation error of a live evolution.

The central empirical result of this work is a separation of the calibrated tolerance into an exponent and a
prefactor. On smooth synthetic fields and on the real slice, the residual-to-error coupling shares essentially
the same power-law exponent, which is close to 2, while the prefactor differs by a factor of order several
hundred in relative error. Part of this factor reflects a genuine difference in absolute sensitivity, and part
reflects the small electric baseline of the real slice, on which a small absolute corruption becomes a large
relative error. Interestingly, the gap is not removed when the residual is placed in a standard dimensionless
form, which indicates that it reflects a solution-dependent and operator-dependent sensitivity rather than
merely a choice of raw units. It is also important to stress what this paper does not claim. It does not introduce a new Weyl implementation,
and it does not present a new inventory of gravitational regimes, since existing invariant characterisation implementations and their applications have begun to establish such an atlas. Rather, it provides the reliability layer
beneath such an atlas, i.e., a means of attaching to every mechanism label a calibrated statement of the
constraint quality required.

\section{Reliability-gated diagnostics}
\label{sec:framework}

We begin by defining the diagnostic vector and the gate that acts on it. A cosmological solution on a spatial
slice is summarised, following the work presented in~\cite{ugail}, by the scalar vector,
\begin{equation}
\Idiag = \big(\,I_\rho,\ I_\theta,\ I_\sigma,\ I_E,\ I_B,\ I_\Ham,\ I_\Mom \,\big).
\label{eq:X}
\end{equation}
The first five components measure the matter inhomogeneity, the expansion inhomogeneity, the shear magnitude,
and the electric and magnetic Weyl energies. The final two components, i.e., the Hamiltonian residual
$I_\Ham$ and the momentum residual $I_\Mom$, are reliability axes rather than physical structure axes, and
they are the subject of this work. Throughout the numerical experiments, the fundamental observer is the
future-directed unit normal to the slice, $u^{a}=n^{a}$, so that all the contractions below are spatial. The
signature is $(-+++)$, and the extrinsic-curvature sign convention is the one for which an expanding Friedmann
slice has $K<0$ and $\theta=-K>0$.

We normalise the two Weyl components by a common curvature scale built from the expansion $\theta=-K$,
\begin{equation}
I_E = \frac{\avg{E_{ij}E^{ij}}}{K_D}, \quad
I_B = \frac{\avg{B_{ij}B^{ij}}}{K_D}, \quad
K_D = \avg{\theta^{4}} + \epsilon_K,
\label{eq:IEIB}
\end{equation}
where $\avg{\cdot}$ denotes the spatial-domain average and $\epsilon_K$ is a small regulator that keeps the
normalisation finite. In the present experiments, it is a numerical floor many orders of magnitude below
$\avg{\theta^{4}}$, and the results are insensitive to its value, since every slice used for calibration has
$\avg{\theta^{4}}$ of order unity or larger. Because $E_{ij}E^{ij}$ and $\theta^{4}$ carry the same dimension,
$I_E$ and $I_B$ are dimensionless, and they remain comparable across solutions of very different absolute
scale. Note that the present calibration is developed for expanding cosmological slices with a non-negligible
$\avg{\theta^{4}}$. On a maximal or a bouncing slice, $\avg{\theta^{4}}$ approaches zero, and the regulator, or
a different curvature scale, becomes essential. This is one reason such slices are unsuitable as calibration
anchors. A classifier reads a mechanism label from $\Idiag$. It returns an electric-Weyl-dominated label when
$I_E$ dominates and $I_B$ lies below a departure floor, a magnetic-Weyl-dominated label in the opposite case,
and a silent label when $I_B$ is negligible. Here, silent is used in the operational sense that $B_{ij}=0$, or
equivalently that $I_B$ lies below the magnetic departure floor, for the normal congruence.

The final two components of Eq.~\eqref{eq:X} are the Einstein constraint residuals. We write the ADM
Hamiltonian and momentum constraints as,
\begin{align}
\Ham &= R + K^{2} - K_{ij}K^{ij} - 2\kappa\rho - 2\Lambda, \label{eq:Ham}\\
\Mom^{i} &= D_{j}\!\left(K^{ij} - \gamma^{ij}K\right) - \kappa\, j^{i}, \label{eq:Mom}
\end{align}
where $R$ is the spatial Ricci scalar, $\gamma_{ij}$ the induced metric, $D_i$ its covariant derivative,
$\kappa=8\pi G$, and $(\rho,j^{i})$ are the energy and momentum densities measured by the normal observer. We
define the reliability axes as norms of these residuals. For the Hamiltonian residual, we reduce the local
scalar $|\Ham|$ over the domain, and for the momentum residual, we reduce the local magnitude
$|\Mom|=(\gamma_{ij}\Mom^{i}\Mom^{j})^{1/2}$ over the domain. Two reductions appear below. The controlled
experiments in Sections~\ref{sec:mechanism} to \ref{sec:calibration} use the peak amplitude, i.e.,
$I_\Ham=\max|\Ham|$ and $I_\Mom=\max|\Mom|$, which gives the strictest local statement. The dimensionless
analysis in Section~\ref{sec:convention} uses the root-mean-square, since that reduction admits the fractional
normalisation defined there. Both reductions increase monotonically with the perturbation amplitude. Hence, the form of the calibration is unaffected by the choice, while the numerical tolerance value depends on it and is reported for the reduction used.

The gate itself is a predicate on the diagnostic vector. Because the tolerance is sector-specific, the gate
takes the form,
\begin{equation}
I_\Ham < \tau_\Ham(\varepsilon_{E,\star}) \quad\text{and}\quad I_\Mom < \tau_\Mom(\varepsilon_{B,\star}),
\label{eq:gate}
\end{equation}
where $\tau_\Ham$ certifies a target accuracy $\varepsilon_{E,\star}$ in the electric axis and $\tau_\Mom$
certifies a target accuracy $\varepsilon_{B,\star}$ in the magnetic axis. A classification is admitted only
where the relevant residual falls below its calibrated tolerance, and it is otherwise recorded as
reliability-limited. The fixed choice $\tau_\Ham=\tau_\Mom=10^{-3}$ adopted in~\cite{ugail}, together
with a departure floor of $10^{-4}$ and a weak-field tolerance of $10^{-2}$, is recovered as a reference
special case. In the remainder of this paper, we measure the tolerance that certifies a stated accuracy, and
we examine whether a single value can serve every solution. As we show, it cannot.

\section{Mechanism and calibration in controlled solutions}
\label{sec:mechanism}

We first establish the behaviour of the gate on constructions in which the reference values are known
analytically, so that the response of the gate and the form of the residual-to-error coupling can be read
against a controlled reference. We build a matched electric and magnetic pair on an expanding background, and
we inject a constraint-violating perturbation of amplitude $\eta$ while we track the induced error in the
dominant Weyl axis alongside the reliability residual. The two constructions follow standard scalar and tensor
perturbation practice, together with the interpretation of the constraints in the $3+1$
setting~\cite{bardeen,kodama,mukhanov,shibata,bssn}. Their explicit form, the noise model, and the grid and
derivative scheme are given in Appendix~\ref{app:methods}.

\begin{figure}
\centering
\includegraphics[width=0.62\textwidth]{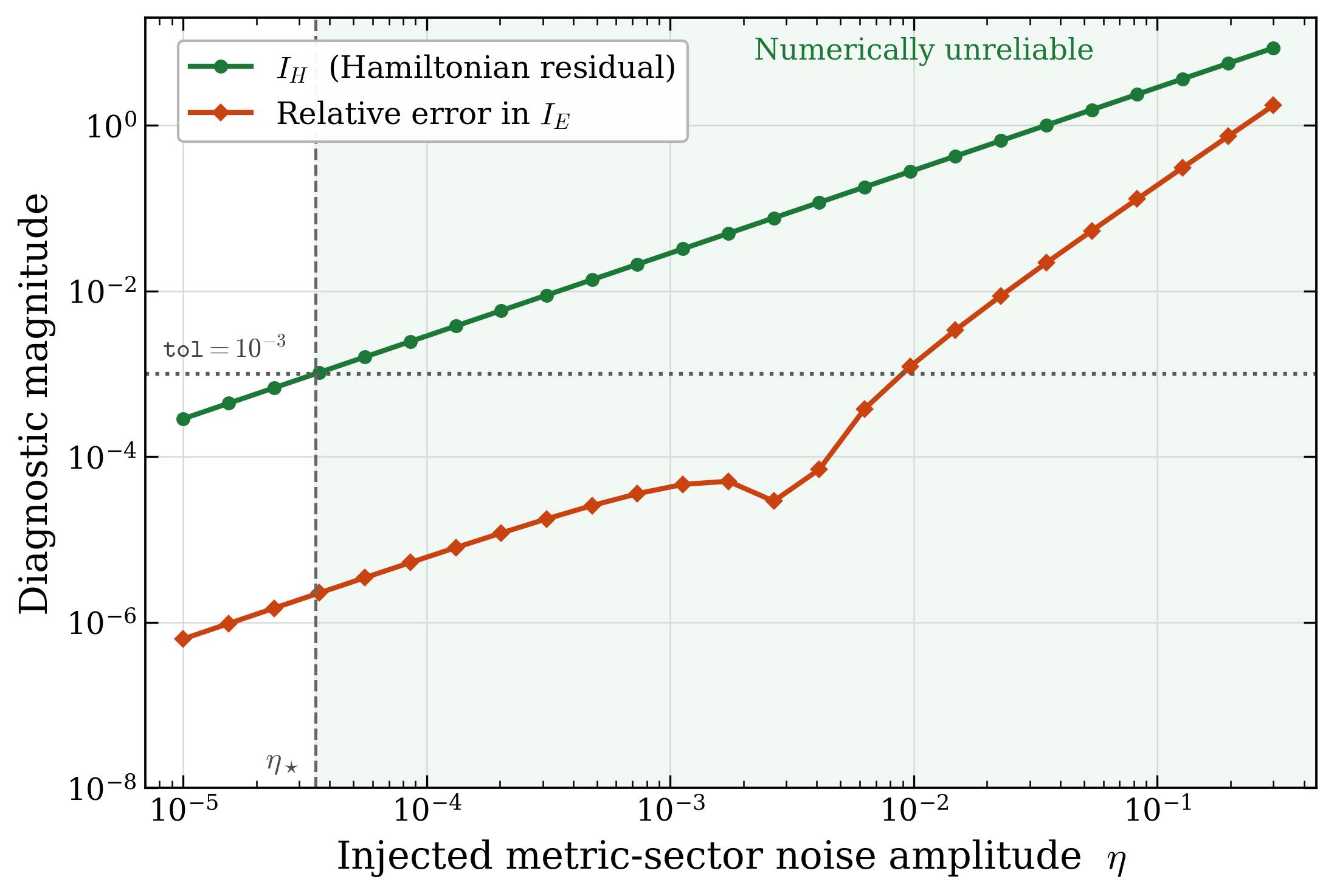}
\caption{The electric-sector reliability gate on the synthetic scalar construction. The Hamiltonian residual
$I_\Ham$ and the relative error in the electric Weyl axis $I_E$ both rise with the injected metric-sector
noise amplitude $\eta$. At the fixed reference tolerance, the gate trips at $\eta_\star$, marked by the dashed
line, while the error in $I_E$ remains near one part in $10^{6}$.}
\label{fig:gate-e}
\end{figure}

\begin{figure}
\centering
\includegraphics[width=0.62\textwidth]{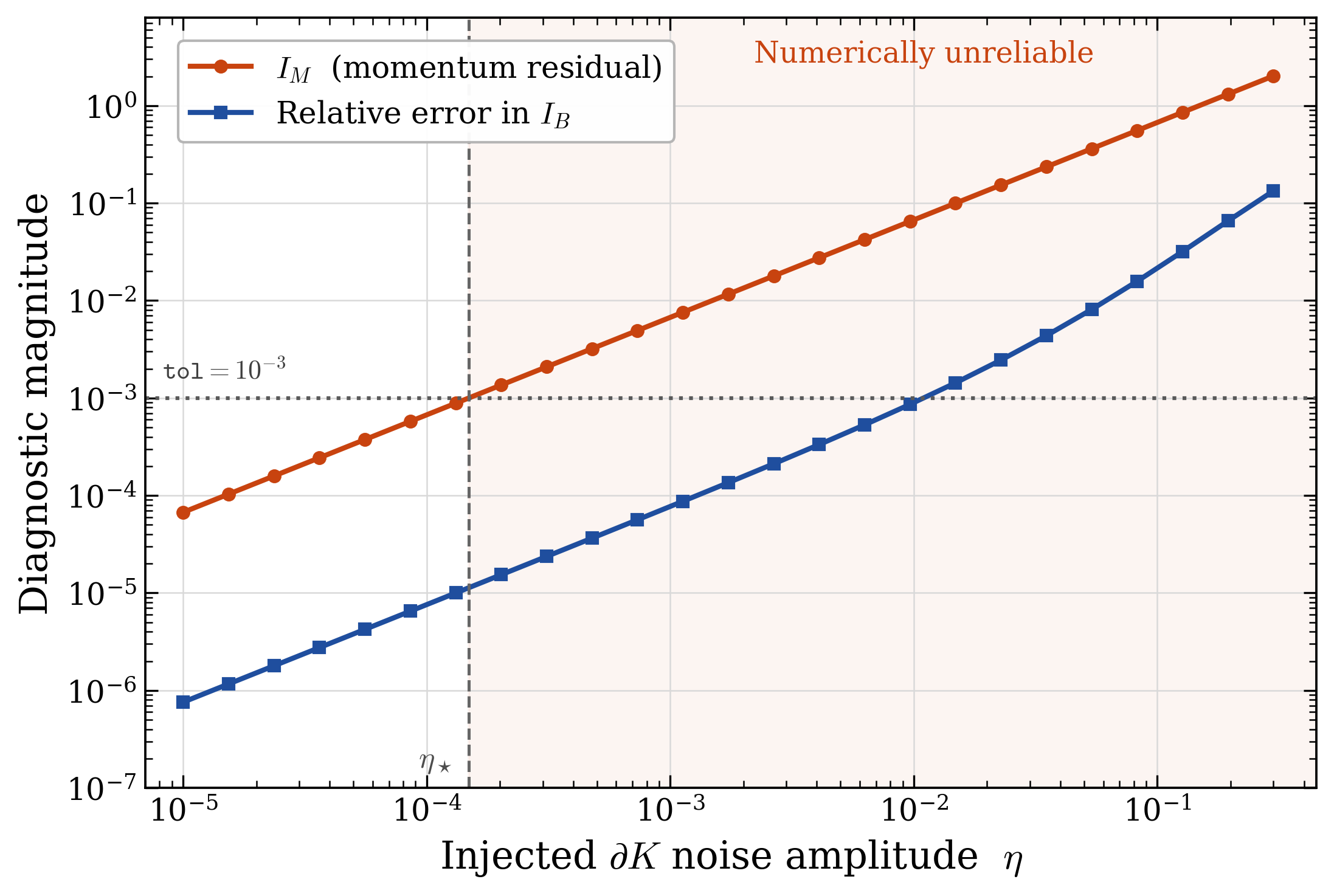}
\caption{The magnetic-sector reliability gate on the synthetic transverse-traceless construction. The
momentum residual $I_\Mom$ and the relative error in the magnetic Weyl axis $I_B$ both rise with the injected
extrinsic-curvature noise amplitude $\eta$. The gate trips at $\eta_\star$ while the error in $I_B$ remains
near one part in $10^{5}$.}
\label{fig:gate-m}
\end{figure}

In the electric sector, a scalar potential $\Phi$ on an expanding slice generates a purely electric
configuration, in which $E_{ij}$ is the trace-free Hessian of $\Phi$ and $B_{ij}$ vanishes by construction.
The classifier returns an electric-Weyl-dominated label with $I_E=0.3048$ and a vanishing clean Hamiltonian
residual. When we inject metric-sector noise, that residual rises, and at the fixed reference tolerance, the
gate first trips at $\eta_\star\approx3.5\times10^{-5}$, where the error in $I_E$ is only about
$2.2\times10^{-6}$, as shown in Fig.~\ref{fig:gate-e}. In the magnetic sector, the twin construction places a
transverse-traceless extrinsic-curvature perturbation on the same background. Here, both $B_{ij}$ and the
momentum residual $\Mom^{i}$ are first derivatives of $K_{ij}$, so the magnetic axis is the natural partner of
the momentum gate. The classifier returns a magnetic-Weyl-dominated label with $I_B=0.3752$ and a vanishing
clean momentum residual, and the gate first trips at $\eta_\star\approx1.5\times10^{-4}$, where the error in
$I_B$ is about $1.1\times10^{-5}$, i.e., close to one part in $10^{5}$, as shown in Fig.~\ref{fig:gate-m}.
Thus, the two sectors establish the reliability pairing,
\begin{equation}
I_\Ham \longleftrightarrow I_E, \qquad I_\Mom \longleftrightarrow I_B ,
\label{eq:pairing}
\end{equation}
which is one of the central claims of this work.

At the fixed reference tolerance, the gate is conservative in the sense of false negatives. It records the
classification as reliability-limited at a noise level where the Weyl axis is still accurate to a part in
$10^{5}$ or better. This is a property of the fixed reference threshold rather than of the diagnostic itself,
and it is precisely the reason we replace that threshold by a calibrated tolerance. Obviously, the margin of
four to five orders of magnitude between the point at which the reference gate trips and the point at which
the classification becomes materially wrong is what the calibration recovers.

A near-quadratic power law is expected for these mappings on physical grounds. The Weyl energies are quadratic
contractions of second derivatives of the metric, while the injected residual grows approximately linearly in
the perturbation amplitude $\eta$. Consider a quadratic energy $Q=\avg{E_{ij}E^{ij}}$ perturbed by
$E\to E+\delta E$. This contains both a mixed clean-noise term $2\avg{E\,\delta E}$ and a pure noise-squared
term $\avg{\delta E^{2}}$. After the domain and ensemble averaging, the noise-squared term dominates over the
fitted range, which gives exponents near two. The residual mixed term, the finite-difference operators, and
the baseline amplitude of the clean field explain the departures from exact quadratic scaling, and they can
produce exponents below two, as we see in the magnetic case. Thus, this argument turns the subsequent
power-law fit from a purely empirical device into a physically motivated model, and it explains why the fitted
exponents cluster near two rather than taking a value fixed by convention.

\begin{figure}
\centering
\includegraphics[width=0.62\textwidth]{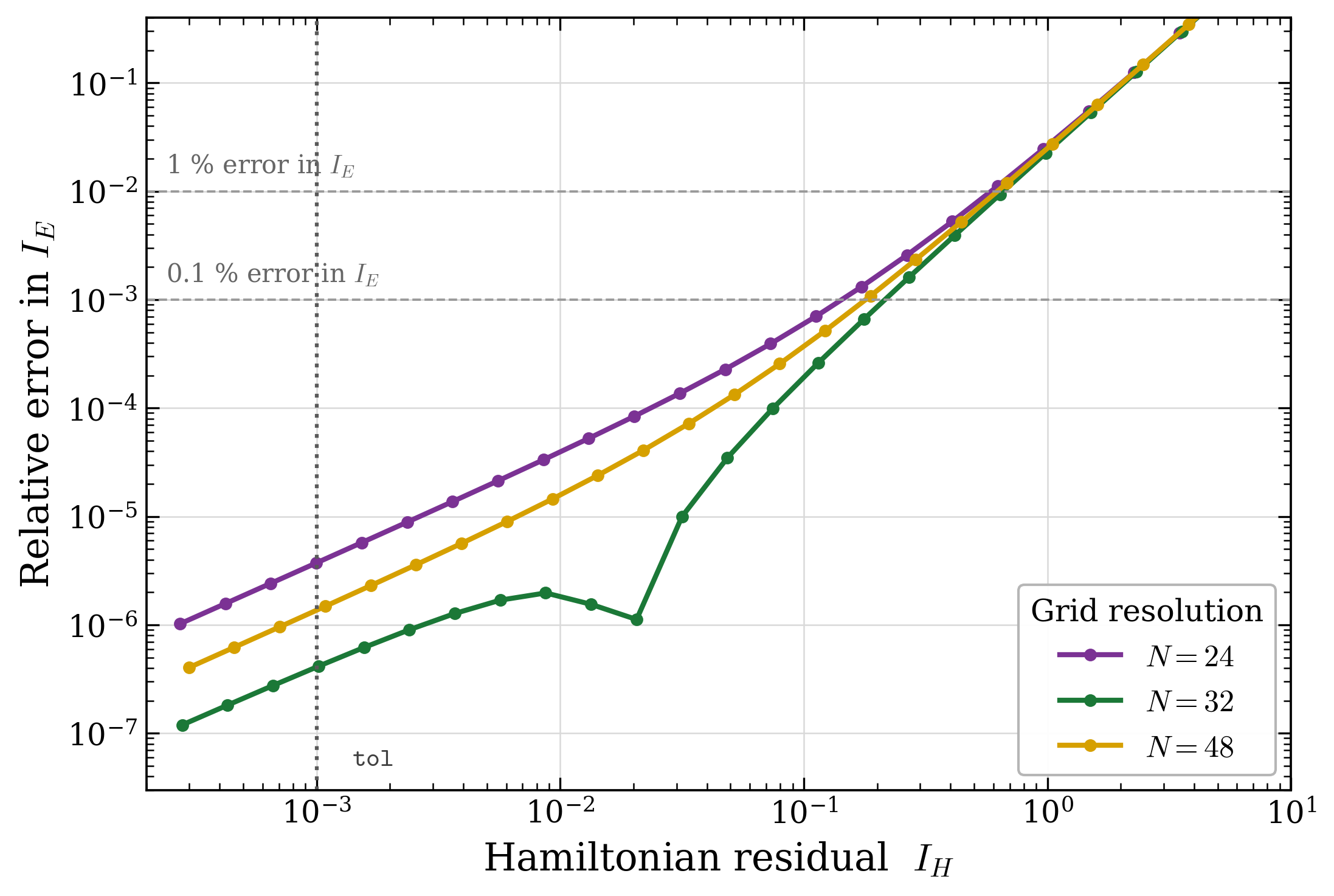}
\caption{Residual-to-error calibration for the electric coupling from $I_\Ham$ to $I_E$ at grid resolutions
$N=24$, $32$, and $48$. The relation is monotonic and well behaved but shifts with resolution, so the calibration must be carried out at the resolution of the data being classified.}
\label{fig:cal-e}
\end{figure}

\begin{figure}
\centering
\includegraphics[width=0.62\textwidth]{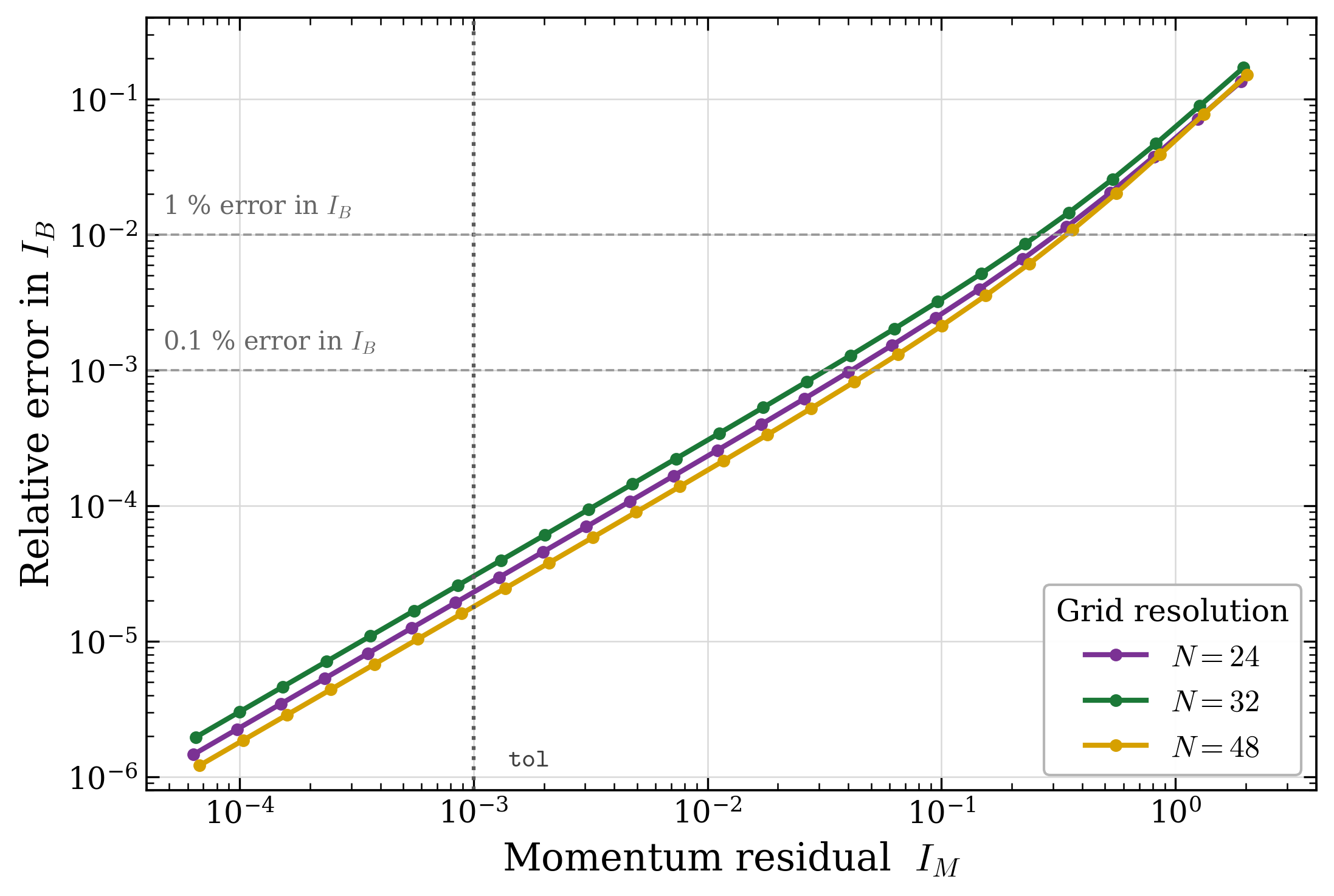}
\caption{Residual-to-error calibration for the magnetic coupling from $I_\Mom$ to $I_B$ at grid resolutions
$N=24$, $32$, and $48$. As in the electric sector, the relation is monotonic and resolution dependent.}
\label{fig:cal-m}
\end{figure}

The residual-to-error map is monotonic and well described by this power law. However, it is not a fixed
property of the sector alone. It shifts with the grid resolution $N$ at which the fields and the constraints
are discretised, as shown in Figs.~\ref{fig:cal-e} and \ref{fig:cal-m} for $N=24$, $32$, and $48$. The trend
is smooth, and it does not alter the pairing of Eq.~\eqref{eq:pairing}. Nevertheless, it implies that the map,
and hence any tolerance derived from it, must be measured at the working resolution rather than transported
from another. This is a first instance, at the level of discretisation, of the non-portability that becomes
far more pronounced on real data in Section~\ref{sec:realdata}. In short, no single value applies across
resolutions, and still less across solutions.

\section{Non-portability of a fixed tolerance on real data}
\label{sec:realdata}

We now replace the synthetic constructions by a real inhomogeneous cosmological initial-data slice, and we
hand the computation of the constraints and the Weyl scalars to an independent implementation. The data is a perturbed Friedmann slice of the silent class. It is generated from relativistic curvature perturbations of the kind
used to set non-Gaussian and relativistic initial conditions, and applied in numerical relativity structure
formation studies~\cite{bruniIC,munozbruni}. All the constraint residuals, the electric and magnetic Weyl
scalars, and the normalising curvature are evaluated with \textsc{aurel}~\cite{aurel}, which is the successor
to the invariant characterisation implementation of~\cite{ebweyl}. The gauge and frame dependence inherent in such
diagnostics has been analysed in the cosmological setting~\cite{ijjas,aurrek}. The reliability of the present
gate is meaningful only if it holds under the same machinery a practitioner would employ, and this is why we
perform the check against that machinery rather than through an independent re-derivation. For the synthetic
constructions, the reference values are analytic. For the real slice, there is no closed-form benchmark, and we
measure the error relative to the clean \textsc{aurel} evaluation of the unperturbed initial data rather than
to an exact solution.

\begin{figure}[t]
\centering
\includegraphics[width=0.94\textwidth]{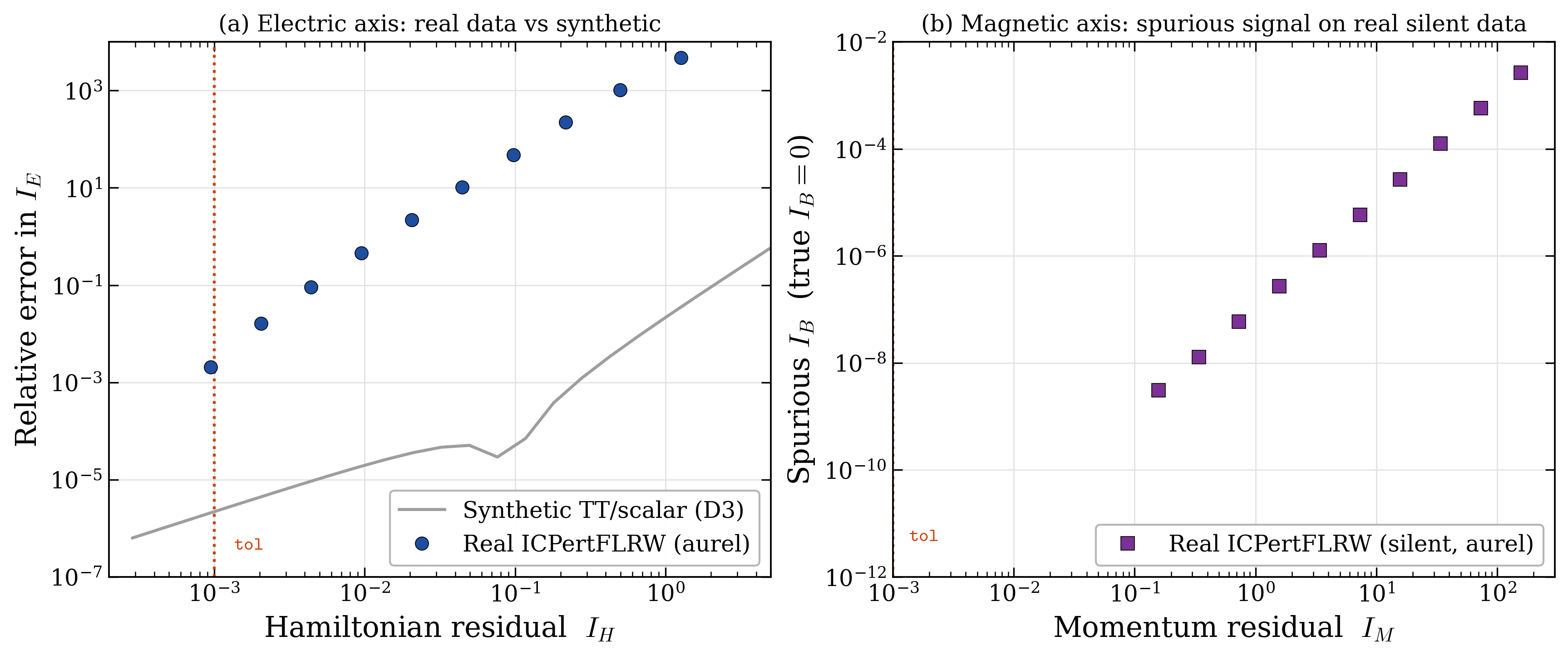}
\caption{Real perturbed-Friedmann initial data with constraints and Weyl scalars computed by \textsc{aurel}.
Panel (a) compares the electric-axis error against the synthetic result and shows that the real slice reaches
a given relative error at a smaller Hamiltonian residual. Panel (b) shows that momentum-sector noise generates
a magnetic Weyl part on a slice that is silent at the level of $10^{-10}$, which is the false positive the
gate must suppress.}
\label{fig:realdata}
\end{figure}

The clean configuration has expansion $\theta=-K=2.0$, and it is silent to numerical precision, with
$I_B\approx3.7\times10^{-10}$ and a small but non-zero electric axis $I_E\approx3.5\times10^{-6}$. We inject
constraint-violating perturbations into the real spatial metric in the electric sector and into the extrinsic
curvature in the magnetic sector, after which \textsc{aurel} recomputes the residuals and the Weyl scalars.
Two features emerge, both of which are shown in Fig.~\ref{fig:realdata}. First, the qualitative coupling holds,
i.e., a larger constraint residual produces a larger Weyl-axis error, monotonically, exactly as in the
controlled case. Hence, the gate mechanism transfers. Second, and more importantly, the calibration does not
transfer. At the fixed reference tolerance of $10^{-3}$, the electric-axis error on the real data is about $0.27$\%, whereas the same tolerance on the synthetic field of Section~\ref{sec:mechanism} corresponds to an
error of order $10^{-4}$\%. Thus, a single constant certifies markedly different accuracies on the two
solutions.

The silent character of the data makes this concern more acute, and it ties the result directly to the
physical reading of $B_{ij}$. Because the clean slice has $I_B\approx3.7\times10^{-10}$, the momentum-sector noise generates a magnetic Weyl part where the clean configuration has essentially none. If we leave this
spurious $I_B$ uncorrected, it would register as genuine gravitomagnetic, and possibly radiative, structure
that is in fact absent. Note that the test probes the generation of a spurious magnetic part on a single
slice, which is the false-positive mode of failure most relevant to classification near the silent boundary.
It does not probe the evolution of a genuine magnetic mechanism, for which a silent slice need not remain
silent under evolution. The significance of a non-zero magnetic Weyl part in cosmological
settings~\cite{clifton} makes it essential that the momentum reliability axis accompany any physical reading
of $I_B$.

\section{Constraint-calibrated tolerances}
\label{sec:calibration}

The non-portability of a fixed tolerance motivates the central construction of this work, i.e., we measure the
coupling, and we set the tolerance from it. For a given solution and a given Weyl axis, we sweep the injected
noise, we record the constraint residual $I$ and the induced relative error $\varepsilon$ in that axis, and we
fit a power law,
\begin{equation}
\varepsilon = C\, I^{p}.
\label{eq:powerlaw}
\end{equation}
For a target accuracy $\varepsilon_\star$, the certifying tolerance is the inverse of this relation,
\begin{equation}
I_\star(\varepsilon_\star) = \left(\frac{\varepsilon_\star}{C}\right)^{1/p} ,
\label{eq:invert}
\end{equation}
which is the residual below which the classification is accurate to $\varepsilon_\star$ under the calibrated
model. Thus, we replace the hand-set constant with a quantity that is measured rather than assumed, and which
carries an explicit, if model-dependent, accuracy statement. The principle that the outputs of numerical
relativity must be assessed against a stated accuracy requirement, rather than against a universal threshold,
is well established in the analysis of waveform error for gravitational-wave data~\cite{jan}. Here, the
calibrated tolerance applies the same principle to cosmological Weyl diagnostics.

We perform the fits over the interval of relative error between $10^{-3}$ and $1$. This interval brackets the
$0.1$ to $10$\% target band, it excludes the smallest residuals where a numerical floor dominates, and
it excludes the largest residuals that lie outside the perturbative corruption regime.
Table~\ref{tab:tolerances} reports, for each mapping, the number of fitted points, the fitted interval in the
residual, the coefficient and the exponent with the $95$\% confidence interval on the exponent, the
coefficient of determination, and the calibrated 1\%  tolerance with its confidence interval. The
synthetic mappings use 8 points, and they yield tightly constrained exponents. The real electric mapping
uses 4 points within this band, and its exponent carries a correspondingly wider interval, i.e., from about
$1.9$ to $2.8$. However, the point estimate is stable, since extending the interval to include the next point
shifts the exponent from $2.33$ to $2.25$, which lies within the interval. The confidence intervals on the one
per cent tolerances propagate the joint uncertainty in the coefficient and the exponent by a Monte Carlo
sampling of the fitted covariance. Note that we perform the fits on the ensemble means. Hence, these intervals
quantify the regression uncertainty around the fitted mean curve, and they do not include the systematic
uncertainty from the choice of noise model. For completeness, at the stricter and the looser targets, the same
fits give a synthetic electric tolerance of $0.24$ at $0.1$\% and $2.12$ at $10$\%, and a real
electric tolerance of $6.6\times10^{-4}$ and $4.7\times10^{-3}$.

\begin{table}[t]
\centering
\caption{Constraint-calibrated tolerances from Eqs.~\eqref{eq:powerlaw} and \eqref{eq:invert} in raw
residual units, with fit windows, point counts, and $95$ per cent confidence intervals. The exponents cluster
near two, in line with the quadratic argument of Section~\ref{sec:mechanism}, while the one per cent tolerances differ by orders of magnitude across solutions. The stricter and looser targets follow from the same fits.}
\label{tab:tolerances}
\begin{tabular*}{\textwidth}{@{\extracolsep{\fill}}lcccccc@{}}
\toprule
Mapping & $n$ & fit range in $I$ & $C$ & $p$ ($95\%$ CI) & $R^{2}$ & $I_\star(1\%)$ ($95\%$ CI) \\
\midrule
$I_\Mom\!\to\!I_B$, syn  & $8$ & $[0.10,\,2.0]$ & $0.042$ & $1.52\ (1.44,1.60)$ & $0.997$ & $0.39\ (0.37,0.41)$ \\
$I_\Ham\!\to\!I_E$, syn  & $8$ & $[0.28,\,5.6]$ & $0.020$ & $2.12\ (2.06,2.17)$ & $0.999$ & $0.71\ (0.70,0.73)$ \\
$I_\Ham\!\to\!I_E$, real & $4$ & $[9.5\!\times\!10^{-4},\,9.5\!\times\!10^{-3}]$ & $2.6\!\times\!10^{4}$ & $2.33\ (1.90,2.76)$ & $0.996$ & $1.8\!\times\!10^{-3}\ (1.6,1.9)\!\times\!10^{-3}$ \\
\bottomrule
\end{tabular*}
\end{table}

The procedure generalises into a reporting standard for Weyl-based classification. First, we compute the
diagnostic vector on each domain. Then, for the solution family and the working resolution, we carry out a
short controlled residual sweep in each sector, and we record the induced Weyl-axis error as a function of the
residual. Next, we fit a power law to that relation, and we retain its coefficients. We then select a target
accuracy, and we invert the fitted relation to give the tolerance for each sector. Finally, we accept the
electric axis at a point where the Hamiltonian residual falls below its tolerance, and we accept the magnetic
axis where the momentum residual falls below its tolerance. 

\begin{figure}
\centering
\includegraphics[width=0.62\textwidth]{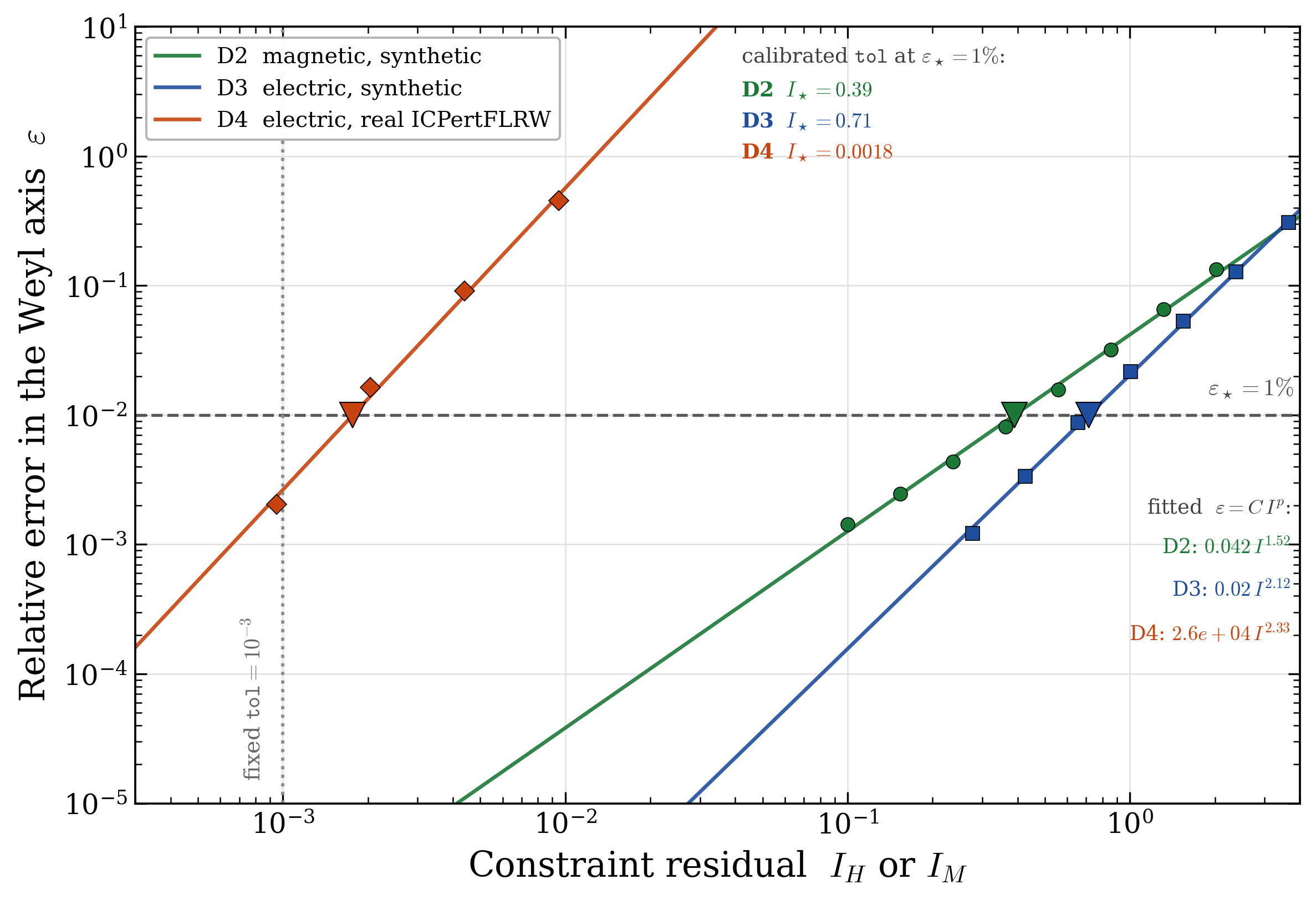}
\caption{Constraint-calibrated tolerances in raw residual units. The residual-to-error map is fitted for each
solution and inverted at a one per cent target to give $I_\star$. The calibrated tolerances span orders of
magnitude between the synthetic fields and the real slice.}
\label{fig:calibration}
\end{figure}

Any domain that fails either test is recorded as
reliability-limited rather than assigned a mechanism. Obviously, the controlled sweep is inexpensive relative
to the computation it certifies, since it reuses the same diagnostic evaluation on a small number of perturbed
copies of a single slice. The resulting calibration may be reused across nearby snapshots or closely related
members of a solution family, but only after a check that the residual-to-error map remains stable. A
materially different regime, i.e., one with altered geometry, resolution, or gauge behaviour, warrants a fresh
sweep. In practice, a sweep of ten to twenty noise amplitudes spanning the target-error band, averaged over a
small number of realisations, is sufficient to constrain the two fit parameters, at a cost of a few diagnostic
evaluations rather than a fraction of the underlying computation. The calibrated tolerances are shown in
Fig.~\ref{fig:calibration}.

\section{Convention robustness and the origin of the gap}
\label{sec:convention}

\begin{figure}
\centering
\includegraphics[width=0.62\textwidth]{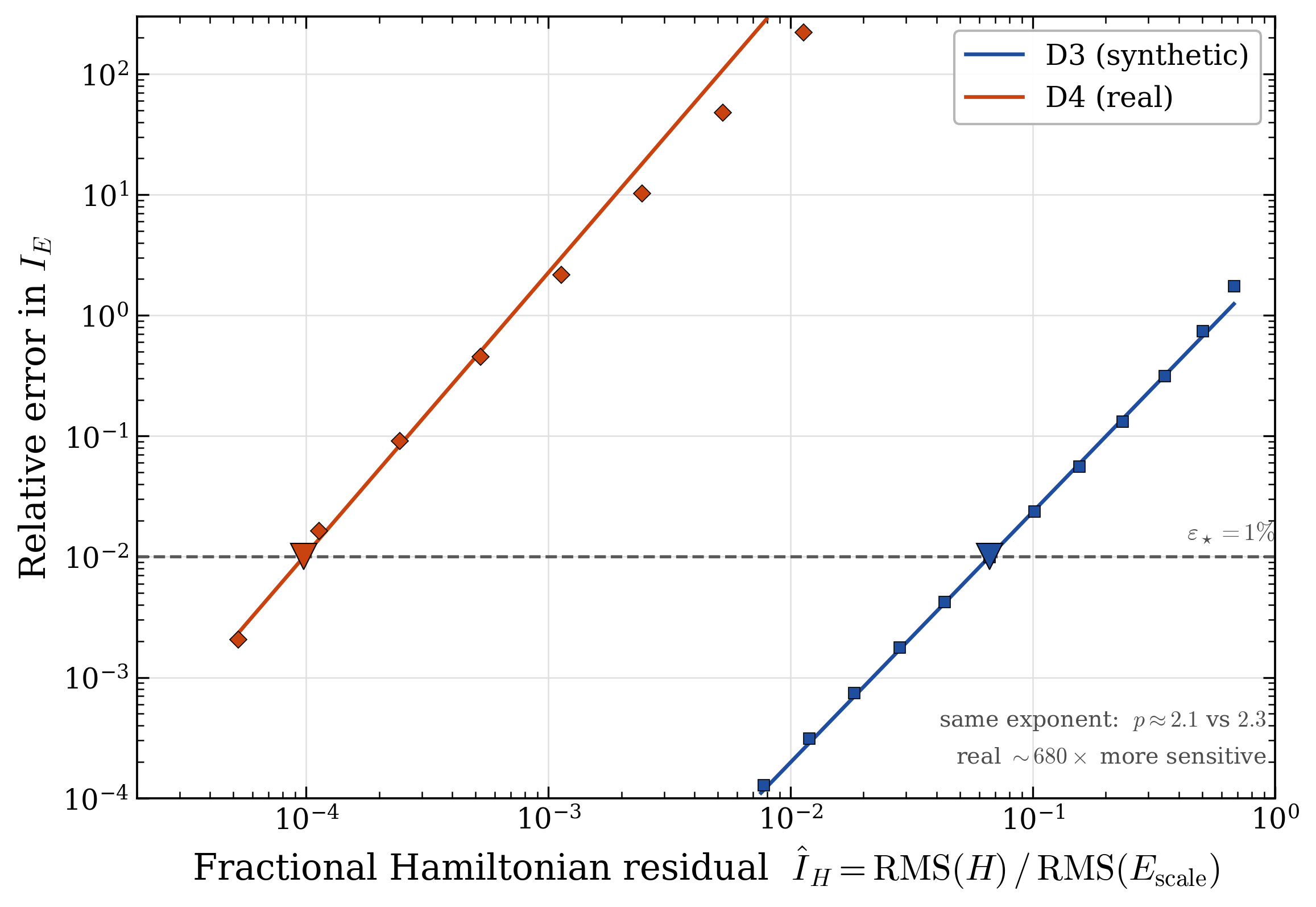}
\caption{Convention robustness of the calibrated tolerance. Expressing the Hamiltonian residual as the
dimensionless fractional residual $\hat I_\Ham=\rms(\Ham)/\rms(\Escale)$ does not close the gap between
synthetic and real data. The two mappings share the same exponent, close to $2.1$ against $2.3$, and differ
in prefactor.}
\label{fig:convention}
\end{figure}

A natural objection is that the spread in $I_\star$ might be an artefact of the residual units. The synthetic
and the real residuals are compared at different absolute scales, and the raw peak norm is not obviously
commensurable across solutions of very different curvature. To address this, we re-express the Hamiltonian
residual in the dimensionless fractional form standard in numerical relativity, in which we normalise the
constraint violation by the root-mean-square of the terms that compose the constraint,
\begin{align}
\hat{I}_\Ham &= \frac{\rms(\Ham)}{\rms(\Escale)}, \nonumber\\
\Escale &= \sqrt{R^{2} + K^{4} + \big(K_{ij}K^{ij}\big)^{2} + (2\kappa\rho)^{2} + (2\Lambda)^{2}} .
\label{eq:frac}
\end{align}
This is the normalisation carried out natively by the analysis code. The numerator and the denominator scale in
the same way under a change of overall scale, so that $\hat I_\Ham$ is genuinely scale-free and directly
comparable between the synthetic and the real electric sectors. When we recompute the residual-to-error map,
refit Eq.~\eqref{eq:powerlaw}, and re-invert Eq.~\eqref{eq:invert}, we obtain Table~\ref{tab:frac} and
Fig.~\ref{fig:convention}.

\begin{table}[t]
\centering
\caption{Calibrated tolerances in the dimensionless fractional residual $\hat I_\Ham$ of Eq.~\eqref{eq:frac}
for the electric sector, with point counts and $95$ per cent confidence intervals. The gap between synthetic
and real data does not close under dimensionless normalisation.}
\label{tab:frac}
\begin{tabular*}{\textwidth}{@{\extracolsep{\fill}}lccccc@{}}
\toprule
Mapping (fractional) & $n$ & $C$ & $p$ ($95\%$ CI) & $R^{2}$ & $\hat I_\star(1\%)$ ($95\%$ CI) \\
\midrule
$I_\Ham\!\to\!I_E$, syn  & $8$ & $2.8$ & $2.08\ (2.03,2.12)$ & $0.999$ & $6.6\!\times\!10^{-2}\ (6.5,6.8)\!\times\!10^{-2}$ \\
$I_\Ham\!\to\!I_E$, real & $4$ & $2.3\!\times\!10^{7}$ & $2.33\ (1.90,2.76)$ & $0.996$ & $9.8\!\times\!10^{-5}\ (9.0,11)\!\times\!10^{-5}$ \\
\bottomrule
\end{tabular*}
\end{table}

The gap does not close. In fact, it widens slightly, and it reaches a ratio of about $680$ at the one per cent
target, against about $400$ in raw units. Furthermore, it is robust to the definition of $\Escale$, since
restricting Eq.~\eqref{eq:frac} to its geometric terms changes the ratio only to about $560$. The two
exponents are nearly equal, i.e., close to $2.1$ for the synthetic field and $2.3$ for the real data. Hence,
the coupling follows essentially the same power law in both cases, and the two curves on logarithmic axes are
nearly parallel. What differs is the prefactor.

The origin of this prefactor difference deserves care, because the ratio is expressed in relative error and
the clean electric axis of the real slice is small, of order $3.5\times10^{-6}$, so a given absolute
corruption produces a large relative error. To separate the two effects, we repeat the comparison in absolute
terms, using the induced change $\Delta I_E$ in place of the relative error. In absolute terms, the real slice
reaches a fixed corruption at a residual smaller than the synthetic field by a factor of about five, rather
than several hundred. Furthermore, at a fixed fractional residual, the induced $\Delta I_E$ on the real slice
exceeds that on the synthetic field by a factor of about twenty. We did not examine the resolution dependence
of this absolute factor, and we leave it to future work. Both statements are informative. The factor of about
five measures a genuinely greater absolute sensitivity of the real geometry, which is not removed by the
dimensionless normalisation, and it is therefore not an artefact of the residual convention. The larger
relative factor measures the classification hazard that a small absolute corruption represents when it sits on
a low-amplitude electric baseline close to a mechanism boundary. Note that the relative measure is the
classification-relevant one, since a mechanism label compares the axis magnitudes against one another and
against the departure floors, and the low-amplitude electric baseline is precisely the regime in which a false
positive is most consequential. Thus, the reliability gap reflects a solution-dependent sensitivity in both an
absolute and a classification-relevant sense, rather than merely a choice of raw units.

\section{Discussion and limitations}
\label{sec:discussion}

The framework discussed in~\cite{ugail} separates the physical structure axes from the reliability axes, and the
present results show that this separation is not cosmetic. The residuals $I_\Ham$ and $I_\Mom$ determine
whether a Weyl classification can be trusted, and the calibrated gate turns them into an explicit, if
model-dependent, accuracy statement. In particular, because a silent slice can acquire a spurious $I_B$ under
momentum-sector corruption, as we showed in Section~\ref{sec:realdata}, the magnetic Weyl energy should not be
interpreted as gravitomagnetic or radiative structure in the absence of the accompanying momentum gate.

The most important limitation concerns the nature of the injected constraint violation. In a live evolution,
the constraint-violating modes are not random. They grow with a formulation-dependent propagation and gauge
couplings, for example in the conformal or the Z4 systems, and they carry a characteristic spatial structure.
Our construction instead injects a smooth low-wavenumber random perturbation into the metric and the extrinsic
curvature. Hence, it measures the sensitivity of the slice-level diagnostics to a generic constraint
perturbation rather than reproducing the structure of an evolution driven violation. Nevertheless, the
construction remains informative for two reasons. First, it isolates the residual-to-error coupling of the
diagnostics themselves, independently of any particular evolution scheme. Second, the low wavenumber content
to which the Weyl energies are most sensitive is the part least likely to be suppressed by constraint damping.
A calibration driven by an evolution generated violation, together with its dependence on the evolution
formulation and on the constraint damping parameters, is a natural and important extension, and we leave it to
future work.

Three further qualifications highlight the real-data anchor. First, the independence we claim is of the residual
and Weyl computation, which is performed throughout by \textsc{aurel}, and not of the noise source. Hence, the
result is a verification on an independent implementation rather than a comparison against an alternative or an exact
solution. Second, the magnetic-sector anchor sits on silent data, so it probes the manufacture of a spurious
$I_B$ from a baseline near $10^{-10}$ rather than the corruption of a pre-existing magnetic mechanism. Third,
the synthetic and the real solutions differ in absolute scale, so the comparison in
Sections~\ref{sec:calibration} and \ref{sec:convention} is properly one of the shape and the slope of the
coupling, made rigorous by the dimensionless normalisation of Eq.~\eqref{eq:frac} and by the absolute-error
analysis, rather than a numerical overlay of raw residual values.

Here, we present the convention-robustness analysis for the electric sector alone, and this is for a specific
reason. The magnetic and momentum fractional residual is degenerate for the transverse-traceless construction
in Section~\ref{sec:mechanism}, because its clean momentum-constraint constituent terms vanish identically, so
that the analogue of $\Escale$ tends to zero and the fractional ratio is ill-defined. Hence, a dimensionless
magnetic check requires a real, non-silent solution that carries a non-zero momentum-constraint scale, e.g., a
slice containing gravitational-wave content. Supplying the magnetic sector with such an anchor would complete
the matrix of electric against magnetic and synthetic against real in a fully dimensionless form, and the
dimensionless robustness established here should be regarded as an electric-sector result until that check is
performed.

The calibration inherits the dependencies of the underlying discretisation, including the resolution, as shown
in Figs.~\ref{fig:cal-e} and \ref{fig:cal-m}, together with the finite-difference stencil, the gauge, the
observer, any smoothing, and the choice of domain. A full convergence study of the calibration with
resolution, and of its dependence on the evolution formulation, is beyond the present scope, and we note it as
future work. This is a further reason we advocate the calibration per solution family and resolution, rather
than tabulating it as a universal value. A similar lesson is familiar from geometric discretisation more
broadly. Curvature-sensitive numerical schemes, such as biharmonic subdivision on Riemannian
manifolds~\cite{ugail2026biharmonic}, show that the behaviour of a smooth construction can depend on the
underlying geometric operator and on the curvature of the space in which it acts. Hence, it is plausible that
the calibration measured here depends not only on the resolution but also on the discrete operator itself.
Likewise, nonlinear and group-valued geometric discretisations can exhibit regularity effects that are not
apparent from a single refinement step~\cite{ugail2026heisenberg}, and adaptive geometric operators change
their behaviour across Euclidean, spherical, and hyperbolic settings~\cite{ugail2026tension}, i.e., the local
geometry can affect smoothness and fidelity diagnostics. We offer these observations as analogies rather than
as evidence about Weyl tensors, but they reinforce the case for calibrating per operator and per geometry
rather than assuming portability.

The atlas of local gravitational regimes has begun to be mapped by
existing invariant characterisation implementations, and their applications~\cite{ebweyl,munozbruni}, and the
computations here are performed by their successor~\cite{aurel}. The contribution of the present work is
orthogonal to that inventory. It is the reliability layer that states how good the constraints must be before
a given cell label is trustworthy, together with the procedure that calibrates that threshold for the solution
at hand. Natural extensions include non-spherical exact benchmarks and the wider family of exact inhomogeneous
cosmologies as controlled test beds~\cite{tolman,bondi,szekeres,bolejko}, together with the weak-field
relativistic simulation approaches against which the diagnostics can be cross-checked~\cite{adamek,aurrek}.

\section{Conclusions}
\label{sec:conclusions}

In this work, we have turned the constraint residuals of a Weyl-diagnostic framework from monitored
by-products into sector-specific, calibrated reliability gates for numerical relativity cosmology. Our matched
controlled constructions establish the natural pairings of the Hamiltonian residual with the electric Weyl
axis and of the momentum residual with the magnetic Weyl axis, and at the fixed reference tolerance the gate
is conservative in the sense of false negatives. A real perturbed-Friedmann initial-data slice, with all the
residual and Weyl computations performed by an independent implementation, confirms the same qualitative couplings,
while it shows that the quantitative tolerances are not portable. 

Our calibration procedure, which fits a
power law and inverts it to obtain a tolerance for a stated accuracy, replaces a hand-set threshold with a
measured, if model-dependent, accuracy statement. The central finding we report here is that the residual-to-error coupling
shares a common exponent across the synthetic and the real regimes, but a prefactor larger by a factor of
order several hundred in relative error on the real slice, of which a factor of about 5 is a genuine
absolute sensitivity, and the remainder is the small electric baseline of the real slice. Furthermore, this gap
survives the standard dimensionless normalisation. Hence, a single fixed tolerance cannot certify a consistent
accuracy for both regimes, and the reliability tolerance of a Weyl atlas should be calibrated for each solution
and resolution. What results is a map of the geometric structure of a spacetime, annotated cell by cell with a
calibrated statement of its own epistemic reliability. 

Finally, the extension of the dimensionless
magnetic-sector check to a real solution carrying gravitational-wave content, and the calibration driven by a
live evolution rather than an injected perturbation, are the natural next steps.

\section*{Acknowledgments}
The author would like to thank the Centre for Visual Computing and Intelligent Systems at the University of
Bradford for providing the computing resources used in this work.

\section*{Data availability}
The complete experimental code, including the diagnostics, the noise sweeps, and the calibration fits, is
openly available at \url{https://github.com/ugail/Cosmology-Weyl-Reliability}. The real-data computations used
the publicly available \textsc{aurel} package~\cite{aurel}.

\section*{Funding}
No specific funding was received for this work.

\appendix

\section{Construction, noise model, and fitting}
\label{app:methods}

The synthetic experiments use a periodic cubic grid of side $2\pi$ with $N$ points per dimension and a uniform
spacing $\Delta x=2\pi/N$, with $N=32$ unless a resolution is stated. We compute the spatial derivatives with
second-order central finite differences, i.e., first derivatives for the momentum and magnetic quantities and
second derivatives for the electric quantities. The expanding background has $\theta=3H$ with $H=0.5$, so that
$\theta=1.5$.

For the electric sector, the scalar potential is,
\begin{equation}
\Phi = A\left(\sin x + 0.6\sin(2y+0.3) + 0.4\cos(z+x)\right),
\end{equation}
with $A=0.8$, taken with zero spatial mean. The electric Weyl tensor is the trace-free Hessian of $\Phi$, and
$I_E$ is the domain average of its square normalised by $\avg{\theta^{4}}$. We define the clean Hamiltonian
residual by subtracting a fixed source so that the unperturbed field satisfies the constraint, and we evaluate
it as the peak of $|\nabla^{2}\Phi - 3H^{2}\Phi - \text{source}|$.

For the magnetic sector, the extrinsic curvature is $K_{ij} = -H\,\delta_{ij} + \delta K_{ij}$, where the
transverse-traceless perturbation has $\delta K_{xx}=-\delta K_{yy}=A\cos(kz)$ and
$\delta K_{xy}=\delta K_{yx}=A\sin(kz)$ with $A=0.5$ and $k=2$. The perturbation is traceless by construction,
and it is transverse because its only spatial dependence is on $z$ while its non-zero components lie in the
$x$ and $y$ directions, so that $\partial_{i}\delta K^{ij}=0$. The magnetic Weyl tensor is the symmetrised
curl of $K_{ij}$, and the momentum residual is $\Mom^{i}=\partial_{j}K^{ji}-\partial^{i}K$, which we reduce by
its peak magnitude.

We inject the constraint violation as a smooth random field. We pass a unit-variance Gaussian white field
through a Gaussian low-pass filter in Fourier space with cutoff wavenumber $k_c=3$, and we renormalise it to
unit standard deviation. In the electric sector, we displace the potential by $\eta A$ times this field. In
the magnetic and the real-data sectors, we form a symmetric tensor by placing an independent filtered field,
scaled by $\eta$ times the root-mean-square of the field being perturbed, in each independent component. The
synthetic sweeps average over ten random seeds, and the reported $I_\Ham^{\text{std}}$ and
$I_\Mom^{\text{std}}$ are the standard deviations of the residual over this ensemble. The real-data sweep uses
a single seed, because each \textsc{aurel} recomputation is comparatively expensive, and repeating the
real-data calibration over several seeds is a robustness check we leave to future work.

We note that the use of smooth, analytically controlled fields as computational test beds follows a broader
tradition of PDE-based geometric modelling, in which harmonic and biharmonic operators generate smooth,
boundary-controlled geometric fields~\cite{monterde2004,monterde2006}, fourth-order elliptic formulations
underpin practical geometric design~\cite{ugail2011pde}, and the approach as a whole has been
surveyed in~\cite{gonzalez2008}. The synthetic constructions above are in the same spirit, i.e., smooth fields
with a known structure serve as controlled references against which a numerical diagnostic can be calibrated.

The real-data anchor is a perturbed Friedmann slice of the silent class, evaluated with \textsc{aurel} on a
grid of $N=48$ with box side $L=1821$ at time $t=1$, with curvature-perturbation amplitude $A\approx0.0114$,
all in the dimensionless code units of the initial-data construction. The metric perturbation drives the
electric sector and the extrinsic-curvature perturbation drives the magnetic sector, and \textsc{aurel}
computes the residuals, the electric and magnetic Weyl scalars, the fractional residual of Eq.~\eqref{eq:frac}, and the normalising curvature.

We perform the fits of $\varepsilon=C\,I^{p}$ by ordinary least squares in logarithmic variables over the
interval of relative error between $10^{-3}$ and $1$. The confidence intervals on the exponent are the
standard $95$ per cent intervals of the regression, and we obtain the confidence intervals on $I_\star$ by a
Monte Carlo sampling of the fitted coefficient-exponent covariance, inverting Eq.~\eqref{eq:invert} for each
sample.


\begin{thebibliography}{99}

\bibitem{buch2000}
Buchert, T. (2000). On average properties of inhomogeneous fluids in general relativity: dust cosmologies.
\textit{General Relativity and Gravitation}, 32, 105--125. \url{https://doi.org/10.1023/A:1001800617177}

\bibitem{buch2001}
Buchert, T. (2001). On average properties of inhomogeneous fluids in general relativity: perfect fluid
cosmologies. \textit{General Relativity and Gravitation}, 33, 1381--1405.
\url{https://doi.org/10.1023/A:1012061725841}

\bibitem{greenwald}
Green, S. R., Wald, R. M. (2011). New framework for analyzing the effects of small scale inhomogeneities in
cosmology. \textit{Physical Review D}, 83, 084020. \url{https://doi.org/10.1103/PhysRevD.83.084020}

\bibitem{giblin}
Giblin, J. T., Mertens, J. B., Starkman, G. D. (2016). Departures from the
Friedmann-Lemaitre-Robertson-Walker cosmological model in an inhomogeneous universe: a numerical examination.
\textit{Physical Review Letters}, 116, 251301. \url{https://doi.org/10.1103/PhysRevLett.116.251301}

\bibitem{bentivegna}
Bentivegna, E., Bruni, M. (2016). Effects of nonlinear inhomogeneity on the cosmic expansion with numerical
relativity. \textit{Physical Review Letters}, 116, 251302. \url{https://doi.org/10.1103/PhysRevLett.116.251302}

\bibitem{macph2017}
Macpherson, H. J., Lasky, P. D., Price, D. J. (2017). Inhomogeneous cosmology with numerical relativity.
\textit{Physical Review D}, 95, 064028. \url{https://doi.org/10.1103/PhysRevD.95.064028}

\bibitem{macph2019}
Macpherson, H. J., Price, D. J., Lasky, P. D. (2019). Einstein's universe: cosmological structure formation
in numerical relativity. \textit{Physical Review D}, 99, 063522.
\url{https://doi.org/10.1103/PhysRevD.99.063522}

\bibitem{adamek}
Adamek, J., Daverio, D., Durrer, R., Kunz, M. (2016). gevolution: a cosmological N-body code based on general
relativity. \textit{Journal of Cosmology and Astroparticle Physics}, 2016(07), 053.
\url{https://doi.org/10.1088/1475-7516/2016/07/053}

\bibitem{aurrek}
Aurrekoetxea, J. C., Clough, K., Lim, E. A. (2025). Cosmology using numerical relativity.
\textit{Living Reviews in Relativity}, 28, 5. \url{https://doi.org/10.1007/s41114-025-00058-z}

\bibitem{ugail}
Ugail, H. (2026). Geometric phase-space structure in cosmological solutions of Einstein's field equations.
arXiv:2606.17071. \url{https://doi.org/10.48550/arXiv.2606.17071}

\bibitem{buch2020}
Buchert, T., Mourier, P., Roy, X. (2020). On average properties of inhomogeneous fluids in general relativity
III: general fluid cosmologies. \textit{General Relativity and Gravitation}, 52, 27.
\url{https://doi.org/10.1007/s10714-020-02670-6}

\bibitem{ijjas}
Ijjas, A. (2023). Gauge/frame invariant variables for the numerical relativity study of cosmological
spacetimes. \textit{Journal of Cosmology and Astroparticle Physics}, 2023(06), 061.
\url{https://doi.org/10.1088/1475-7516/2023/06/061}

\bibitem{maartens}
Maartens, R., Bassett, B. A. (1998). Gravito-electromagnetism. \textit{Classical and Quantum Gravity}, 15,
705--717. \url{https://doi.org/10.1088/0264-9381/15/3/018}

\bibitem{costa}
Costa, L. F. O., Nat\'ario, J. (2014). Gravito-electromagnetic analogies. \textit{General Relativity and
Gravitation}, 46, 1792. \url{https://doi.org/10.1007/s10714-014-1792-1}

\bibitem{clifton}
Clifton, T., Gregoris, D., Rosquist, K. (2017). The magnetic part of the Weyl tensor, and the expansion of
discrete universes. \textit{General Relativity and Gravitation}, 49, 30.
\url{https://doi.org/10.1007/s10714-017-2192-0}

\bibitem{owen}
Owen, R., Brink, J., Chen, Y., Kaplan, J. D., Lovelace, G., Matthews, K. D., Nichols, D. A., Scheel, M. A.,
Zhang, F., Zimmerman, A., Thorne, K. S. (2011). Frame-dragging vortexes and tidal tendexes attached to
colliding black holes: visualizing the curvature of spacetime. \textit{Physical Review Letters}, 106, 151101.
\url{https://doi.org/10.1103/PhysRevLett.106.151101}

\bibitem{munozbruni}
Mu\~noz, R. L., Bruni, M. (2023). Structure formation and quasispherical collapse from initial curvature
perturbations with numerical relativity simulations. \textit{Physical Review D}, 107, 123536.
\url{https://doi.org/10.1103/PhysRevD.107.123536}

\bibitem{ebweyl}
Mu\~noz, R. L., Bruni, M. (2023). EBWeyl: a code to invariantly characterize numerical spacetimes.
\textit{Classical and Quantum Gravity}, 40, 135010. \url{https://doi.org/10.1088/1361-6382/acd6cf}

\bibitem{shibata}
Shibata, M., Nakamura, T. (1995). Evolution of three-dimensional gravitational waves: harmonic slicing case.
\textit{Physical Review D}, 52, 5428--5444. \url{https://doi.org/10.1103/PhysRevD.52.5428}

\bibitem{bssn}
Baumgarte, T. W., Shapiro, S. L. (1998). On the numerical integration of Einstein's field equations.
\textit{Physical Review D}, 59, 024007. \url{https://doi.org/10.1103/PhysRevD.59.024007}

\bibitem{alcubierre}
Alcubierre, M., Br\"ugmann, B., Diener, P., Koppitz, M., Pollney, D., Seidel, E., Takahashi, R. (2003). Gauge
conditions for long-term numerical black hole evolutions without excision. \textit{Physical Review D}, 67,
084023. \url{https://doi.org/10.1103/PhysRevD.67.084023}

\bibitem{gundlach}
Gundlach, C., Mart\'in-Garc\'ia, J. M., Calabrese, G., Hinder, I. (2005). Constraint damping in the Z4
formulation and harmonic gauge. \textit{Classical and Quantum Gravity}, 22, 3767--3773.
\url{https://doi.org/10.1088/0264-9381/22/17/025}

\bibitem{alic}
Alic, D., Bona-Casas, C., Bona, C., Rezzolla, L., Palenzuela, C. (2012). Conformal and covariant formulation
of the Z4 system with constraint-violation damping. \textit{Physical Review D}, 85, 064040.
\url{https://doi.org/10.1103/PhysRevD.85.064040}

\bibitem{aurel}
Mu\~noz, R. L., et al. (2026). aurel: automated relativistic diagnostics for numerical spacetimes.
arXiv:2602.00155. \url{https://doi.org/10.48550/arXiv.2602.00155}

\bibitem{bardeen}
Bardeen, J. M. (1980). Gauge-invariant cosmological perturbations. \textit{Physical Review D}, 22,
1882--1905. \url{https://doi.org/10.1103/PhysRevD.22.1882}

\bibitem{kodama}
Kodama, H., Sasaki, M. (1984). Cosmological perturbation theory. \textit{Progress of Theoretical Physics
Supplement}, 78, 1--166. \url{https://doi.org/10.1143/PTPS.78.1}

\bibitem{mukhanov}
Mukhanov, V. F., Feldman, H. A., Brandenberger, R. H. (1992). Theory of cosmological perturbations.
\textit{Physics Reports}, 215, 203--333. \url{https://doi.org/10.1016/0370-1573(92)90044-Z}

\bibitem{bruniIC}
Bruni, M., Hidalgo, J. C., Meures, N., Wands, D. (2014). Non-Gaussian initial conditions in $\Lambda$CDM:
Newtonian, relativistic, and primordial contributions. \textit{The Astrophysical Journal}, 785, 2.
\url{https://doi.org/10.1088/0004-637X/785/1/2}

\bibitem{jan}
Jan, A., Ferguson, D., Lange, J., Shoemaker, D., Zimmerman, A. (2024). Accuracy limitations of existing
numerical relativity waveforms on the data analysis of current and future ground-based detectors.
\textit{Physical Review D}, 110, 024023. \url{https://doi.org/10.1103/PhysRevD.110.024023}

\bibitem{ugail2026biharmonic}
Ugail, H., Howard, N. (2026). Biharmonic subdivision on Riemannian manifolds. arXiv:2604.13083.
\url{https://doi.org/10.48550/arXiv.2604.13083}

\bibitem{ugail2026heisenberg}
Ugail, H., Carriazo, A. (2026). A Heisenberg subdivision scheme with central smoothness loss.
arXiv:2607.05446. \url{https://doi.org/10.48550/arXiv.2607.05446}

\bibitem{ugail2026tension}
Ugail, H., Howard, N. (2026). A neural tension operator for curve subdivision across constant curvature
geometries. arXiv:2603.28937. \url{https://doi.org/10.48550/arXiv.2603.28937}

\bibitem{tolman}
Tolman, R. C. (1934). Effect of inhomogeneity on cosmological models. \textit{Proceedings of the National
Academy of Sciences USA}, 20, 169--176. \url{https://doi.org/10.1073/pnas.20.3.169}

\bibitem{bondi}
Bondi, H. (1947). Spherically symmetrical models in general relativity. \textit{Monthly Notices of the Royal
Astronomical Society}, 107, 410--425. \url{https://doi.org/10.1093/mnras/107.5-6.410}

\bibitem{szekeres}
Szekeres, P. (1975). A class of inhomogeneous cosmological models. \textit{Communications in Mathematical
Physics}, 41, 55--64. \url{https://doi.org/10.1007/BF01608547}

\bibitem{bolejko}
Bolejko, K., C\'el\'erier, M.-N., Krasi\'nski, A. (2011). Inhomogeneous cosmological models: exact solutions
and their applications. \textit{Classical and Quantum Gravity}, 28, 164002.
\url{https://doi.org/10.1088/0264-9381/28/16/164002}

\bibitem{monterde2004}
Monterde, J., Ugail, H. (2004). On harmonic and biharmonic B\'ezier surfaces. \textit{Computer Aided
Geometric Design}, 21(7), 697--715. \url{https://doi.org/10.1016/j.cagd.2004.07.003}

\bibitem{monterde2006}
Monterde, J., Ugail, H. (2006). A general 4th-order PDE method to generate B\'ezier surfaces from the
boundary. \textit{Computer Aided Geometric Design}, 23, 208--225.
\url{https://doi.org/10.1016/j.cagd.2005.09.001}

\bibitem{ugail2011pde}
Ugail, H. (2011). \textit{Partial Differential Equations for Geometric Design}. Springer, London.
\url{https://doi.org/10.1007/978-0-85729-784-6}

\bibitem{gonzalez2008}
Gonz\'alez Castro, G., Ugail, H., Willis, P., Palmer, I. (2008). A survey of partial differential equations
in geometric design. \textit{The Visual Computer}, 24, 213--225.
\url{https://doi.org/10.1007/s00371-007-0190-z}

\end{thebibliography}
\end{document}